\documentclass[runningheads]{llncs}
\usepackage[T1]{fontenc}

\usepackage{graphicx} 
\usepackage{colortbl} 
\usepackage[table]{xcolor} 
\usepackage{booktabs} 
\usepackage[most]{tcolorbox} 
\usepackage{subfig}
\usepackage{hyperref}
\usepackage{multirow}
\usepackage{balance} 
\usepackage{stfloats}
\usepackage{xcolor}
\usepackage{amssymb}
\usepackage{listings}
\usepackage{colortbl}
\usepackage[capitalise]{cleveref}
\let\llncssubparagraph\subparagraph
\let\subparagraph\paragraph
\usepackage[compact]{titlesec}
\let\subparagraph\llncssubparagraph

\usepackage{xcolor}

\newcommand{\student}[2]{%
  \textcolor{blue!50}{\emph{``#1''}}%
  ~{\footnotesize   (ID~#2)}%
}

\begin{document}

\title{Humor in Software Testing Education}

\author{Isabella Graßl \inst{1}\orcidID{0000-0001-5522-7737} \and
Benoit Baudry \inst{2}\orcidID{0000-0002-4015-4640}}
\authorrunning{Graßl and Baudry}
%
\institute{Technical University of Darmstadt, Germany \and
Université de Montréal, Canada}



\lstdefinestyle{customjava}{
    language=Java,
    basicstyle=\ttfamily\scriptsize,
    keywordstyle=\color{blue},
    commentstyle=\color{gray},
    stringstyle=\color{orange},
    showstringspaces=false,
    tabsize=4,
    breaklines=true,
    numbers=left,
    numberstyle=\tiny\color{gray},
    xleftmargin=2em,
    frame=single
}

\maketitle

\begin{abstract}
Software testing is often perceived as monotonous, which can negatively influence students’ emotional engagement with testing. While prior work suggests that humor can increase engagement in professional software development contexts, we know little about humor's effect in  software testing education. 
This paper explores how humorous elements in software testing assignments affect students’ emotional engagement, sense of belonging, and creative thinking. We introduced humor in introductory software testing courses at universities in Canada and Germany, and conducted a mixed-methods study with students.
Our results show that humor had a strong positive influence on students’ experiences of software testing. Students perceived testing as more engaging and less monotonous, felt more comfortable and accepted in class, and reported increased creative thinking about testing tasks. These effects were particularly strong for female students, especially with respect to sense of belonging.
Our findings suggest that humor represents a low-threshold pedagogical approach which is benefical for students, and  has the potential to create a more welcoming learning environment. 
\keywords Software testing education \and humor \and creativity

\end{abstract}

\section{Introduction}
Software testing is often perceived as monotonous and frustrating~\cite{mulgund2025,waychal2016}. Such perceptions make it difficult to emotionally engage students with testing and can reduce their motivation to approach testing tasks~\cite{desouzasantos2023a}. This is concerning because software testing is a key part of ensuring sofware quality~\cite{lima2023}. If educators can improve students' feeling about testing, they may become more willing to engage with it, enjoy it, and ultimately write better tests or even consider testing-related career paths~\cite{tantithamthavorn2023a,jokisch2024}.

Several approaches in software testing education aim to increase student engagement and foster positive initial attitudes, such as gamification or active learning~\cite{izu2024,Silvis-Cividjian25}. However, these approaches remain underutilized in many courses~\cite{fasolino2026,arghandwal2022}. One subtle and low-effort approach that has received little attention in this context looks at a fundamental aspect of human interaction: \emph{humor}. Research in psychology, education, and the social sciences shows that humor can support communication, reduce social distance, and encourage creativity, while also easing negative emotions such as stress or tension~\cite{berk2000,bishara2022}.

Despite this relevance, humor remains largely under-researched in software engineering and its education~\cite{hidellaarachchi2025,tiwari2024,arghandwal2022}. Only a small number of studies have examined humor in broader computing education~\cite{barros2017,hu2017}. This gap is interesting given that informal practitioner communities frequently use humor when discussing software~\cite{tiwari2024}. However, we  lack empirical evidence on how humor affects students in formal software testing education.

We address this gap with an exploratory empirical education research study on the affective role of humor in  software testing education. Our objective is to investigate how humor in testing assignments influence students’ perceived emotional engagement, sense of belonging, and creative thinking. 
We integrated humorous elements into software testing lectures and assignments at two universities in Canada and Germany, and conducted a mixed-methods study with 58 students to capture their perceptions and experiences.
To our knowledge, this is one of the first studies to empirically examine humor in software testing education with a focus on emotional and social dimensions.

Our results show that humor had a positive influence on students’ experiences of software testing. The majority of students reported higher engagement, a stronger sense of belonging, and increased creative thinking about testing tasks. These effects were particularly strong for female students, while differences between countries were small.

\emph{Contributions.} (1) We provide first empirical evidence that humor may function as a social-emotional practice in software testing education, improving how students emotionally \emph{and} socially engage with testing tasks.
(2) We show that humor affects students’ sense of belonging and creative thinking, which goes beyond the anticipation of humor \emph{just} makes testing more fun. (3) We identify gender-dependent effects of humor, with female students reporting particularly significant benefits, suggesting humor may be a low-threshold pedagogical approach to support inclusion in software testing education.

\section{Background and Related Work}
Humor is an important element of human interaction, which operates on cognitive, emotional, and social levels, and serves multiple functions in communication~\cite{martin2007}. 
On an emotional level, humor can reduce stress, anxiety, and tension, and help people cope with challenging situations~\cite{berk2000}. These effects are particularly relevant in learning contexts, where fear of failure and performance pressure can hinder participation and engagement.

On a social level, studies show that humor supports communication, strengthens relationships, and helps establish shared understanding~\cite{sanford1984adolescent,crawford1991creativity}. Humor often signals creativity and social awareness, and lower social barriers by easing interaction~\cite{crawford1991creativity}. Such effects may support participation and inclusion, especially for women or adolescents with disorders~\cite{bishara2022,buxbaum2022you}.
At the same time, humor can be perceived differently across contexts as disciplinary and cultural backgrounds shape how people interpret and value humor~\cite{wear2006}. This highlights the need to study humor within specific domains.


\subsection{Humor in Education}
A key challenge in education is to engage students, particularly in subjects often perceived as difficult or boring~\cite{jones2014humor}.
Prior research across disciplines in secondary and higher education shows that humor has social and emotional benefits, e.g. by increasing attention, engagement, reducing tension, and improving communication~\cite{christman2018instructor,hu2017,ashrafzade2025,nieto2022didactic}. 
Studies from mathematics~\cite{bishara2022,amran2022,menezes2021} and other professional training contexts~\cite{chabeli2015humour,ashrafzade2025} report that humor helps students feel more relaxed, improves interaction, and supports understanding of the taught material~\cite{scheel2025}. 
Prior work shows that humorous elements such as puns, visual jokes, or playful language can improve focus and engagement in online courses, since those environments are even more challenging for students to bond with each other~\cite{erdogdu2021a,vogler2019lolsquared}.

In the current higher education discourse, humor is often discussed in \emph{Nature's Career} column as a factor supporting critical thinking, creativity and scientific method~\cite{raman2023tiktok,heidt2023say}, particularly at the graduate level~\cite{baudry2025}. Studies that ask students to actively practice humor, rather than only consume it, suggest that engaging with humor can foster creative thinking and critical thinking skills~\cite{spork2023students}. 


\subsection{Humor in Software Engineering (Education)}


Despite its relevance in general education, humor has received limited attention in software engineering (education). 
This gap is interesting given that software engineering is a socio-technical field: prior work with professional software developers suggests that humor supports engagement with complex problems, improves code readability~\cite{tiwari2024}, and is positively associated with creativity~\cite{khan2022,amjed2016}.

In software engineering education, previous work has not studied the role of humor yet. Instead, related work focuses on fun and gameplay to increase engagement~\cite{PrasetyaLMTBEKM19,fraser2020,jokisch2024}. While effective, these approaches often require substantial effort and resources. 
Prior work has shown that visual and narrative formats, e.g. comics, can improve students’ learning and understanding of software engineering concepts~\cite{barros2017,jwo2015teaching}. Similarly, \_why\_the\_lucky\_stiff wrote a book about Ruby, relying on humor and cartoons, to make programming more appealing \cite{why}.  
Recent work suggests that integrating fun activities in software engineering education improves students understanding of abstract concepts~\cite{Maalej2026OnFF}. 
Our work complements these previous studies as we focus on humor (puns, pranks and comedy references) instead of games and fun activities. Also, we address the emotional and social dimensions of fun in software testing education which is not covered by prior work.

\section{Course Context}\label{sec:course}
We integrated humor into both a software testing lecture and its assignments, aligning humorous elements with core course content, as humor may support positive attitudes when it is meaningfully tied to what students are asked to learn~\cite{scheel2025}.
We provide materials of the course for replication.\footnote{\url{https://figshare.com/s/1677ac2d70e1846cab1a}}

\subsection{Lecture}

The course in Canada is a 3-credit course on software quality and metrics. It covers the key concepts of software testing and analysis at different stages of software development, such as unit testing, test oracle, test quality, static testing, dependency management, security testing or testing in production. Every lecture blends theorertical presentations about quality concepts, with live demos of tools that support these concepts. 

The goal of the course is to introduce fundamental concepts of software testing, as well as the state of the art techniques and tools to automate testing. The key messages is that any testing activity must have a clear intention, must be measurable with respect to an expected behavior and must be automated in order to scale and provide actionable feedback to developers. Most of these concepts are new for the students, yet essential for them to learn as most of them go on the job market at the end of the year when they take this course. 

\begin{figure}[t]
    \centering

    \subfloat[
        Meme about unit testing, used to introduce common pitfalls of code coverage in test suites.
        \label{fig:unit}
    ]{
        \includegraphics[width=0.44\columnwidth]{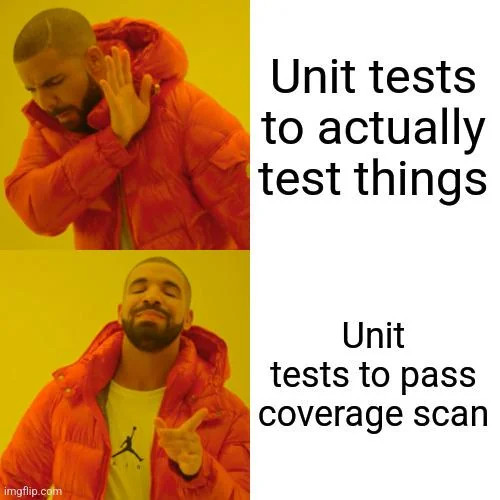}
    }
    \qquad
    \subfloat[
        Example of a \href{https://github.com/lolcommits/lolcommits/wiki/Lolcommits-from-around-the-world\%21}{group commit photo} created with \href{https://lolcommits.com/}{lolcommits}, which illustrates how humor can support engagement and a sense of community in version control.
        \label{fig:lolcommit}
    ]{
        \includegraphics[width=0.44\columnwidth]{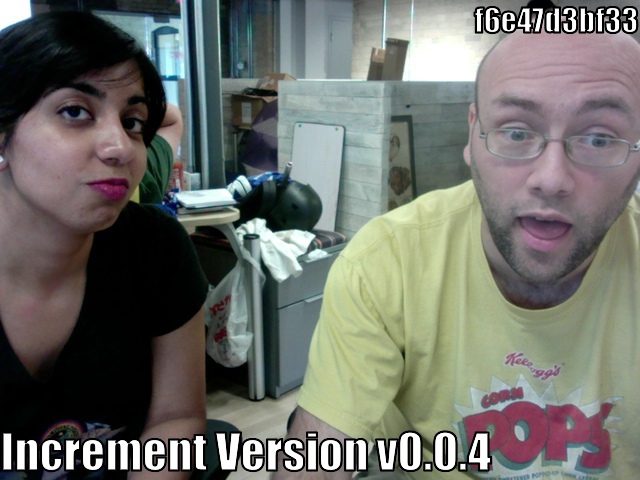}
    }
\end{figure}

In order to increase the engagement of students with these concepts and methods, the course is very interactive, follows a flipped classroom  approach and makes use of humor in several different ways throughout the semester. 
First, the professor uses humor to get the students' attention and pass some concepts that are usually considered as boring or dry. For example, when introducing the key concept of code coverage to determine the quality of unit tests, we use the meme in \autoref{fig:unit}. The image is an appropriate visual cue to discuss the potential caveats of code coverage. The meme's anchor in pop culture facilitates the comprehension and the memorization of this discussion.
Second, the students are given several opportunities to practice humor in software testing, through the assignments.



\subsection{Assignments}
The course has 13 lectures, and the students have to deliver 4 assignments. Assignment \#1 is a presentation performed in person, in class; assignment \#2 is a technical assignments related to unit testing and mutation analysis; assignment \#3 is a technical assignment related to integration testing and mocks; assignment \#4 consists in answering 3 quizzes at different moments in the semester. 

Some students (10\% of the group) do their presentation (assignment \#1) on the theme ``humor and testing'', as part of the lecture that is fully dedicated to this theme. The students who give a presentation must select a topic that touches upon one technique, concept or case study at the intersection of humor and testing. Through these presentations, we discuss the various ways of having humor in test data, e.g. with faker libraries, in comments, e.g. with references to fun novels or TV series, on in commits, e.g. with the systematic use of lolcommits, as illustrated in \autoref{fig:lolcommit}.
The learning objective for this lecture is two-fold. The students who prepare a presentation have to explore the field, choose a specific topic of humor for their presentation and deep-dive into the tool, practice, or concept they choose. All students get an overview of the diversity of tools and practices they can use to introduce humor in their daily testing activities, they learn about the benefits and possible pitfalls of humor for testing.

All the students have to practice humor as part of assignments \#2 and \#3. For assignment \#2, we ask them to use the java-faker library \footnote{\url{https://github.com/DiUS/java-faker}}. This library is useful for testing as it can generate random, realistic test data, such as names, addresses, phone numbers or geographical coordinates. In addition to its utility for testing, jave-faker is also an inspiring library for developers who like combining the rigor of testing with the lightness of humorous references. As such, many developers have extended the library so that it generates names of movie characters, addresses of TV series characters, quotes of famous shows that can be used instead of lorem ipsum, etc. Through assignment \#2, the students have the opportunity to discover this library, both for as a testing utility and as a space for creative expression. 

As part of assignment \#3, the students are asked to modify an existing Github action workflow for integration testing, and the workflow should \footnote{\href{https://youtu.be/dQw4w9WgXcQ}{rickroll}} in case of failure. 
Here the learning objective is two-fold. First, learn how to use a continuous integration workflow such as Github actions, as part of software testing. In particular here, they learn how to break the build if the mutation score the program under test is lower than on the previous commit. Second,  learn how humor can be used responsibly in order to mitigate the frustration of a failing build. In particular, they learn about the various ways of leveraging one the oldest pranks in Internet culture \cite{buchel2012internet}, as part of a continuous pipeline. 

The German course followed the same pedagogical approach and covered the same topics, humor in lectures, faker libraries, rickrolling in CI pipelines, but in a condensed format designed to fit within a single lecture and one assignment.

\section{Methods}
Our aim is to explore how humor influences students perception about testing, focusing on their engagement, sense of belonging, and creative thinking.

\noindent
    \textbf{RQ1:} How do humorous elements in testing influence students’ emotional engagement with software testing? \\
\textbf{RQ2:} How do humorous elements in testing shape students’ sense of belonging?  \\
\textbf{RQ3:} How do humorous elements in testing support students’ creative thinking in software testing?\\

\subsection{Instrument}
We conducted a survey with both closed and open-ended questions based on prior literature. \Cref{tab:survey} shows all survey items related to our RQs. We conducted a pilot with two students from the German university in order to verify the phrasing and order of the question.

\begin{table}[t]
\centering
\caption{The questionnaire for the students.}
\label{tab:survey}
\tiny
\begin{tabular}{p{1cm}p{11cm}}
\toprule
\textbf{Var.} & \textbf{Question} \\
\midrule
\multicolumn{2}{l}{\textbf{Emotional Engagement (RQ1)}} \\ 
\midrule
EM00& My level of engagement in software testing, compared to before the assignment, is now [...] \\
EM01& My level of stress about software testing, compared to before the assignment, is now [...] \\
EM02& The humor in the assignment made testing more enjoyable. \\
EM03& The humor in the assignment made testing more engaging. \\
EM04& The humor in the assignment made testing less monotonous. \\
EM05& The humor in the assignment made testing less frustrating. \\
EM06& The humor in the assignment made me feel less anxious about testing. \\
EM07& The humor in the assignment made testing less intimidating.  \\
EM08 &How did experiencing humor in testing influence your perception of testing? \\
EM09& Would you like to see more humor integrated into more software engineering courses? Why or why not? \\
\midrule
\multicolumn{2}{l}{\textbf{Sense of Belonging (RQ2)}} \\
\midrule
SB00& My feeling of how well I fit into software development, compared to before the assignment, is now [...]\\
SB01& I feel accepted in this class during the assignment.  \\
SB02& I feel comfortable in this class during the assignment.  \\
SB03& I feel supported in this class during the assignment.  \\
SB04& I feel like I belong in this class during the assignment. \\
SB05& I feel the humor about testing made me feel more connected to my classmates.  \\
SB06& I feel the humor made collaboration on testing topics more enjoyable.  \\
SB07& I feel the humor can break down social barriers in the classroom. \\
SB08& The humor used in the assignment can make testing more inclusive.  \\
SB09& How did experiencing humor in testing influence your sense of belonging?  \\
\midrule
\multicolumn{2}{l}{\textbf{Creativity (RQ3)}} \\ 
\midrule
CR00& My level of creativity in software testing, compared to before the assignment, is now [...]  \\
CR01& The humor in the assignment encouraged me to think more creatively about testing.  \\
CR02& The humor helped me consider testing problems from different perspectives. \\
\bottomrule
\end{tabular}
\end{table}

\emph{Demographic Questions. }
We asked about location (Canada or Germany), gender, age, coding experience, prior exposure to humor in software, and students’ reactions to humor in code. 


%

\emph{Emotional Engagement (RQ1). }
We asked ten questions (\emph{EM00--EM09}, \Cref{tab:survey}) about how humor influenced students enjoyment, monotony, frustration, anxiety, and intimidation in testing. We also included an open-ended question about how the humor in the assignments influenced their perception of testing, and whether students wanted more humor in courses and why. These questions build on prior work showing humor can reduce negative emotions and increase engagement~\cite{berk2000,menezes2021,erdogdu2021a}.

\emph{Sense of Belonging (RQ2). }
We asked ten questions (\emph{SB01--SB08}, \Cref{tab:survey}) about acceptance, comfort, support, inclusion, and social connection in class. Students responded using 5-point Likert scales. We also explored whether humor made collaboration more enjoyable or helped break social barriers. Open-ended responses asked how humor influenced students' belonging. Sense of belonging is especially relevant for underrepresented groups~\cite{vorderwulbeke2025belonging,boman2024breaking}.

\emph{Creativity (RQ3). } 
We asked three questions (\emph{CR01--CR03}, \Cref{tab:survey}) about whether humor encouraged creative thinking and considering problems from different perspectives~\cite{erdogdu2021a}. Creativity is an important, yet under-explored, aspect to effective testing, as it helps students think about edge test cases.

\subsection{Data Collection and Participants}
\begin{table}[t]
\centering
\tiny
\caption{Demographics of participants.}
\label{tab:demographics}
\begin{tabular}{p{2.8cm}rrr}
\toprule
Variable & Canada & Germany & Total \\
\midrule
\textbf{Gender} & & & \\
\quad Female & 11 (30.6\%) & 6 (27.3\%) & 17 (29.3\%) \\
\quad Male & 23 (63.9\%) & 16 (72.7\%) & 39 (67.2\%) \\
\quad Non-binary & 1 (2.8\%) & 0 (0.0\%) & 1 (1.7\%) \\
\quad Prefer not to say & 1 (2.8\%) & 0 (0.0\%) & 1 (1.7\%) \\
\midrule
\textbf{Age (n=56)} & & & \\
\quad Mean (SD) & 23.0 (3.3) & 24.9 (7.6) & 23.7 (5.4) \\
\quad Range & 18--35 & 19--56 & 18--56 \\
\midrule
\textbf{Coding Frequency} & & & \\
\quad Almost everyday & 13 (36.1\%) & 8 (36.4\%) & 21 (36.2\%) \\
\quad Once a week & 21 (58.3\%) & 11 (50.0\%) & 32 (55.2\%) \\
\quad Occasionally & 1 (2.8\%) & 3 (13.6\%) & 4 (6.9\%) \\
\quad Rarely & 1 (2.8\%) & 0 (0.0\%) & 1 (1.7\%) \\
\midrule
\textbf{Years of Experience} & & & \\
\quad Mean (SD) & 3.4 (1.7) & 3.0 (1.6) & 3.3 (1.6) \\
\midrule
\textbf{Encountered Humor} & & & \\
\quad Yes & 16 (44.4\%) & 18 (81.8\%) & 34 (58.6\%) \\
\quad No & 20 (55.6\%) & 4 (18.2\%) & 24 (41.4\%) \\
\midrule
\textbf{Reaction to Humor} & & & \\
\quad Would chuckle & 16 (44.4\%) & 13 (59.1\%) & 29 (50.0\%) \\
\quad Want to practice & 19 (52.8\%) & 6 (27.3\%) & 25 (43.1\%) \\
\quad Indifferent & 1 (2.8\%) & 3 (13.6\%) & 4 (6.9\%) \\
\bottomrule
\end{tabular}
\end{table}
We distributed the survey online at the end of the semester. We collected data from Canada ($n=36$) and Germany ($n=22$), which corresponds to response rates of 33\% (CA) and 31\% (DE). Participation was voluntary, anonymous, and not incentivized. According to our universities’ regulations, this study did not require formal ethics approval because it involved anonymous survey data without any intervention. We nonetheless followed institutional ethical guidelines: students could skip questions, responses were fully anonymized, and participation had no impact on grades or course evaluation.

\Cref{tab:demographics} summarizes the demographics of our participants. 
The gender distribution was similar across both countries, with most students identifying as male (67.2\%), followed by female (29.3\%), non-binary (1.7\%), or preferring not to say (1.7\%). This reflects the general gender imbalance in computer science and software engineering.

The average age of students was in the early to mid-20s ($\varnothing 23.7$), with Canadian students slightly younger than German students. Most students (over 90\%) coded regularly, at least once a week, and had moderate programming experience, suggesting they were primarily intermediate-level programmers, typically in their second year of undergraduate studies. 
When asked how they would react to humor in a software project, most responded positively, either chuckling or feeling encouraged to use humor themselves, while very few were indifferent (\Cref{tab:demographics}).

\subsection{Data Analysis}

\subsubsection{Quantitative Analysis}
We analyzed all 5-point Likert-scale responses (\Cref{tab:survey}) using descriptive statistics (mean, percentages). 
To compare groups (gender, country) we used the \emph{Mann-Whitney U} test with a significance level of $\alpha = 0.05$. This non-parametric test is appropriate for our data since they are ordinal and cannot be assumed to follow a normal distribution~\cite{kerby2014simple}. 
To quantify the effect of observed differences, we calculated the rank-biserial correlation ($r_{rb}$) as the effect size measure between a dichotomous variable (\emph{Male/Female, Canada/Germany}) and an ordinal variable (Likert-scale rankings). We interpret effect sizes as small ($r_{rb}=0.10$), medium ($r_{rb}=0.30$), or large ($r_{rb}=0.50$)~\cite{kerby2014simple}. 
We compared students from Canada with students from Germany, as well as female students with male students. We excluded non-binary students from the gender comparison due to small sample size.

\subsubsection{Qualitative Analysis}
We analyzed the open-ended responses (\emph{EM08, EM09, SB09}; \Cref{tab:survey}) using qualitative content analysis following Mayring's approach~\cite{mayring2021qualitative}. This method is well-suited as it enables both inductive category development and systematic coding procedure for relatively brief textual data~\cite{mayring2021qualitative}.

We began with the familiarization phase, reading through all responses multiple times to gain an overall understanding of the data. To establish coding reliability, we used stratified random sampling based on gender and country to ensure representation across our main demographic groups. We randomly selected approximately 25\% of all responses ($n=15$) for independent coding.

Two researchers independently conducted inductive coding on these 15 responses, developing codes that emerged directly from the data rather than applying predefined categories. Given that most responses were relatively brief (one to two sentences), codes were designed to capture the core meaning of each statement. After independent coding, we calculated inter-rater reliability using Krippendorff's $\alpha$~\cite{krippendorff2022}, obtaining $\alpha = 0.92$, which indicates almost perfect agreement~\cite{marzi2024k}. 
Given the high inter-rater reliability, one researcher coded all remaining responses using the coding scheme~\cite{marzi2024k}. We abstracted the initial codes ($n=128$ unique codes) into themes through an iterative process of grouping semantically similar codes, resulting in four main themes and several subthemes (\Cref{tab:themes}). We calculated the frequency of each theme and subtheme as the percentage of participants whose responses included at least one code within that category. We selected representative quotes to illustrate each theme.

\subsection{Threats to Validity}
We considered several threats to the validity of our findings.

\emph{Internal Validity.} 
Students were not randomly assigned to humorous and non-humorous conditions, as all participants experienced humor in their courses. We therefore cannot claim causal effects. Other factors, such as course design or instructor style, may have influenced students’ responses. To reduce this risk, both instructors used comparable assignments, covered similar testing topics, and integrated humor in similar ways. Our data relies on self-reported perceptions, which may differ from objective measures of learning or performance. However, perceptions of engagement, belonging, and creativity are meaningful outcomes in their own way, as they influence students’ motivation and participation.

\emph{External Validity.}
The results may differ in other institutional or cultural contexts. However, we observed similar patterns across both countries, despite differences in prior exposure to humor in software. The courses also enrolled international and culturally diverse student populations, which supports awareness of transferability to similar software engineering education contexts. We encouarge researchers for replication with larger cohorts.

\emph{Construct Validity.}
We measured the influence of humor using self-report items. Other methods, such as behavioral observations or analysis of test artifacts, might reveal different perspectives. However, surveys with data triangulation are appropriate for measuring perceptions and feelings, which were our main focus. We did not differentiate between types or intensity of humor, which may influence outcomes differently.

\emph{Researcher Positionality.} 
The authors (\emph{female German, male Canadian}) are experienced software engineering educators who value humor as pedagogical tool. This perspective may have shaped the study design and interpretation. To mitigate this risk, we grounded our instruments in  literature, used transparent analysis procedures, and reported both positive and critical student responses.

\section{Results}
\begin{table}[t]
\centering
\tiny
\caption{Statistical Comparison of students in Canada (CA) vs Germany (DE). \scriptsize Signficant differences are in \textbf{bold}, $p$-value effect, size $r_{rb}$.}
\label{tab:country_comparison}
\begin{tabular}{p{3.8cm}p{0.7cm}p{0.6cm}>{\raggedleft\arraybackslash}p{0.7cm}>{\raggedleft\arraybackslash}p{0.7cm}}
\toprule
Variable & $Mean$ CA & $Mean$ DE &$p$ & $r_{rb}$ \\
\midrule
\multicolumn{5}{l}{\textbf{Emotional Engagement (RQ1)}} \\ \cmidrule{1-1}
EM00 (level of engagement) & 3.44 & 3.40&  0.617 & 0.073 \\
EM01 (level of stress) & 2.97 & 2.68&  0.211 & 0.186 \\
EM02  (more enjoyable) & 4.05 & 3.90 & 0.340 & 0.139 \\
EM03 (more engaging) & 4.16 & 4.31 & 0.533 & 0.091 \\
EM04 (less monotonous) & 4.30 & 4.13 & 0.378 & 0.128 \\
EM05 (less frustating) & 3.50 & 3.50  & 1.000 & 0.001 \\
EM06 (less anxixious) & 3.52 & 3.59  & 0.785 & 0.040 \\
EM07 (less intimitading) & 3.75 & 3.54 &  0.234 & 0.170 \\

\midrule
\multicolumn{5}{l}{\textbf{Sense of Belonging (RQ2)}} \\ \cmidrule{1-1}
SB00 (level of how well i fit)  & 3.52 & 3.36 & 0.313 & 0.140 \\
SB01 (feeling accepted) & 4.33 & 4.04  & 0.078 & 0.246 \\
SB02 (feeling comfortable) & 4.33 & 4.09  & 0.111 & 0.222 \\
\textbf{SB03 (feeling supported)} & \textbf{4.30} & \textbf{3.90}  & \textbf{0.043} & \textbf{0.298} \\
\textbf{SB04 (feeling belonging)} & \textbf{4.41} & \textbf{3.81}  & \textbf{0.002} & \textbf{0.442} \\
SB05 (classmates) & 3.94 & 3.72 &  0.212 & 0.187 \\
SB06 (collaboration) & 4.27 & 4.00 &  0.086 & 0.245 \\
SB07 (social barriers) & 4.25 & 3.90 &  0.073 & 0.261 \\
SB08 (more inclusive) & 4.08 & 3.86 & 0.210 & 0.181 \\
\midrule
\multicolumn{5}{l}{\textbf{Creativity (RQ3)}} \\ \cmidrule{1-1}
CR00 (level of creativity) & 3.94 & 4.04 &  0.471 & 0.088 \\
CR01 (think creatively) & 4.16 & 4.04 &  0.419 & 0.117 \\
CR02 (perspectives) & 3.80& 3.59 &  0.286 & 0.159 \\
\bottomrule
\end{tabular}
\end{table}
\begin{table}[t]
\centering
\tiny
\caption{Statistical Comparison of female (F) vs male (M) students. \scriptsize Signficant differences are in \textbf{bold}, $p$-value effect, size $r_{rb}$.}
\label{tab:gender_comparison}
\begin{tabular}{p{3.7cm}p{0.7cm}p{0.5cm}>{\raggedleft\arraybackslash}p{1.1cm}>{\raggedleft\arraybackslash}p{0.7cm}}
\toprule
Variable & $Mean$ F & $Mean$ M &$p$ & $r_{rb}$  \\
\midrule
\multicolumn{5}{l}{\textbf{Emotional Engagement (RQ1)}} \\ \cmidrule{1-1}
EM00 (level of engagement) & 3.41 & 3.43 &  0.640 & 0.074 \\
EM01 (level of stress) & 2.94 & 2.84 &  0.834 & 0.035 \\
\textbf{EM02  (more enjoyable)} & \textbf{4.47} & \textbf{3.79} & \textbf{0.002} & \textbf{0.474} \\
\textbf{EM03 (more engaging)} & \textbf{4.70} & \textbf{4.02}  & \textbf{0.001} & \textbf{0.528} \\
\textbf{EM04 (less monotonous)} & \textbf{4.76} & \textbf{4.00}  & \textbf{$<0.001$} & \textbf{0.576} \\
EM05 (less frustating) & 3.70 & 3.46 & 0.283 & 0.170 \\
\textbf{EM06 (less anxixious)} & \textbf{3.94} & \textbf{3.38}  & \textbf{0.015} & \textbf{0.379} \\
\textbf{EM07 (less intimitading)} & \textbf{4.17} & \textbf{3.46}  & \textbf{0.001} & \textbf{0.489} \\
\midrule
\multicolumn{5}{l}{\textbf{Sense of Belonging (RQ2)}} \\ \cmidrule{1-1}
SB00 (level of how well i fit) & 3.47 & 3.46  & 0.815 & 0.036 \\
\textbf{SB01 (feeling accepted)} & \textbf{4.52} & \textbf{4.07}& \textbf{0.013} & \textbf{0.371} \\
\textbf{SB02 (feeling comfortable)} & \textbf{4.64} & \textbf{4.07}  & \textbf{0.002} & \textbf{0.462} \\
\textbf{SB03 (feeling supported)} & \textbf{4.52} & \textbf{4.00}  & \textbf{0.012} & \textbf{0.394} \\
\textbf{SB04 (feeling belonging)} & \textbf{4.52} & \textbf{4.02}  & \textbf{0.010} & \textbf{0.397} \\
\textbf{SB05 (classmates)} & \textbf{4.17} & \textbf{3.71}  & \textbf{0.022} & \textbf{0.368} \\
\textbf{SB06 (collaboration)} & \textbf{4.64} & \textbf{3.97}  & \textbf{$<0.001$} & \textbf{0.531} \\
\textbf{SB07 (social barriers)} & \textbf{4.47} & \textbf{3.94}  & \textbf{0.016} & \textbf{0.376} \\
\textbf{SB08 (more inclusive)} & \textbf{4.52} & \textbf{3.76} & \textbf{$<0.001$} & \textbf{0.558} \\
\midrule
\multicolumn{5}{l}{\textbf{Creativity (RQ3)}} \\ \cmidrule{1-1}
\textbf{CR00 (level of creativity)} & \textbf{4.35} & \textbf{3.82}  & \textbf{$<0.001$} & \textbf{0.460} \\
\textbf{CR01 (think creatively) } & \textbf{4.70} & \textbf{3.87}  & \textbf{$<0.001$} & \textbf{0.569} \\
\textbf{CR02 (perspectives)} & \textbf{4.35} & \textbf{3.46} & \textbf{0.001} & \textbf{0.538} \\
\bottomrule
\end{tabular}
\end{table}
\begin{table}[t]
\centering
\caption{Themes and example codes developed from qualitative content analysis.}
\label{tab:themes}
\tiny
\begin{tabular}{p{4.2cm}p{8cm}}
\toprule
(Sub)Theme & Codes  \\
\midrule
\multicolumn{2}{l}{\textbf{Emotional Relief and Enjoyment} \textit{(65.5\%; EM08, EM09, SB09)}}\\
Enjoyment and Motivation \textit{(44.8\%)}
& Fun, Enjoyable, Motivaed, Pleasant, Nice, Surprising (positive) \\
Lightness and Casualness \textit{(36.2\%)}
& Light, Chill, Relaxing, Comfortable, Less boring, Less serious, Mood \\
Stress and Fear Reduction \textit{(15.5\%)} 
& Less stressful, Worry less, Dedramatize fear of failure, Less intimidating \\
\midrule
\multicolumn{2}{l}{\textbf{Learning and Reframing of Testing} \textit{(41.4\%; EM08, EM09)}} \\
Enhanced Learning  \textit{(19.0\%)}
& Vivid, Focusing, Remember easily, Effective, Approachable, For beginners\\
Reframing Testing as Creative \textit{(17.3\%)}
& Creativity, Novelty, Experiment, Exploring, Less technical, Personality \\
\midrule
\multicolumn{2}{l}{\textbf{Social Connection and Belonging} \textit{(36.2\%; SB09)}} \\
Community Building \textit{(27.6\%)}
& Bonding, Fit into group, Welcoming, Friendly, Informal, Less alone \\
Direct Belonging and Inclusion \textit{(13.8\%)}
& Understood, Confident, Contribution independent of experience, Less strict \\
Communication \textit{(10.4\%)}
& Communication, Topic, Fit into group \\
\midrule
\multicolumn{2}{l}{\textbf{Contextual Limitations and Caution} \textit{(31.0\%; EM08, EM09)}} \\
Context-Dependency \textit{(20.7\%)}
& Careful, Depends on context, Depending on course quality \\
Neutral Response  \textit{(19.0\%)}
& Uncertain, Indifferent, Not fundamental, No impact, Offending, Trolling\\
\bottomrule
\end{tabular}
\end{table}
For each of our three RQs, we first report descriptive results and qualitative findings. We then compare statistical differences between countries and genders.

\subsection{RQ1: Humor on Students' Emotional Engagement}
\begin{figure}
  \centering
    \subfloat[\centering Retrospective perception of emotional engagement.]{{\includegraphics[width=0.8\linewidth]{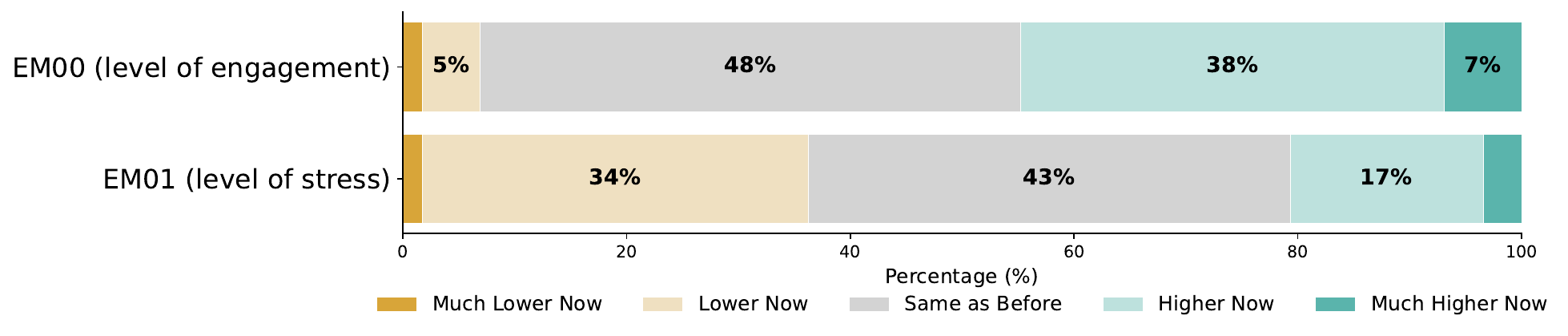} }}%
    \qquad
    \subfloat[\centering Perception of humor on emotional engagement.]{{\includegraphics[width=0.8\linewidth]{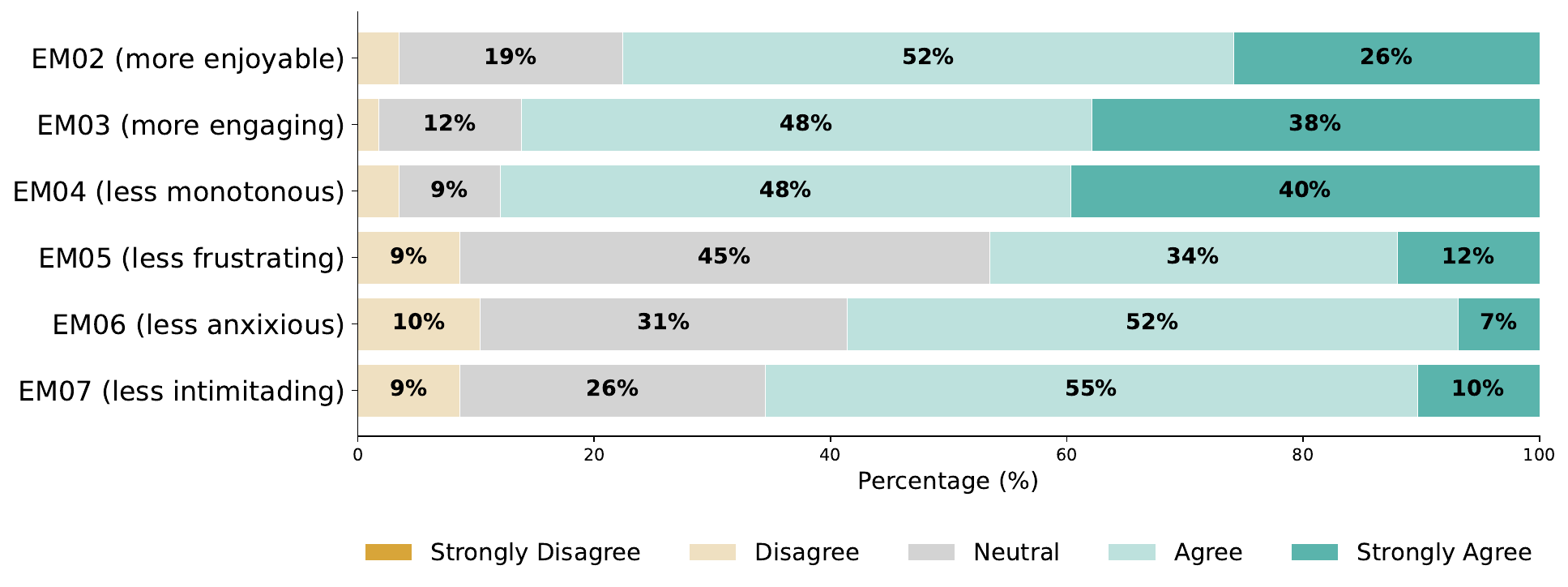} }}%
    \caption{Students' perceived emotional engagement in testing.}%
    \label{fig:emotions}%
\end{figure}
To examine whether humor in testing assignments affects students’ emotional engagement, we asked students about their perceived emotions after the assignment (\emph{EM00--EM09}).

\textbf{Overall Emotional Engagement. } 
\Cref{fig:emotions} shows students' perceived emotional engagement. 
Overall, students agreed that humor had a positive effect on their emotions during the testing assignments. 
Almost half of the students perceived higher enjoyment and engagement in software testing after the assignment (44.8\%,), while about half perceived no change (48.3\%; \emph{EM00}, \Cref{fig:emotions}). Students also reported reduced stress and anxiety, as over one third perceived lower stress compared to before the assignment (36.2\%; \emph{EM01}).

The majority of students reported that the assignment felt less monotonous (87.9\%, \emph{EM04}) and more engaging (86.2\%, \emph{EM03}). Many students also reported that humor made testing more enjoyable (77.6\%, \emph{EM02}). 
We observed moderate effects for emotions related to reducing negative emotions. Students reported that humor made testing less intimidating (65.5\%, \emph{EM07}), reduced anxiety about testing (58.6\%, \emph{EM06}), and reduced frustration (46.6\%, \emph{EM05}).

\textbf{Qualitative Insights.} 
The qualitative analysis (\emph{EM08, EM09}) revealed that students valued humor because it brought a human element to their technical work. As one student put it, humor \student{makes testing much more interesting and funny. Code without emotions feels like code from an LLM}{42-DE-M}.\footnote{The ID followed by a number represents a student from our study from Canada (CA), Germany (DE), female (F), or male (M).}

\emph{Enjoyment and Lightness.} 
According to our students, humor actively created positive experiences. Students described testing as becoming more fun. This emotional shift mattered because it changed how students approached their work. When testing felt lighter and more enjoyable, students seemed more willing to engage with it rather than treating it as a task to complete quickly.

\emph{Stress Reducation.} 
Students described how humor reduced the emotional burden of testing. One student noted that humor \student{made debugging less painful}{02-CA-F}, while another explained that \student{when you’re frustrated with coding, a funny line of code can lighten the mood, bring new motivation, and create a moment of relaxation}{47-DE-M}. These responses suggest that humor helped students manage the frustration that often accompanies debugging and testing.

\emph{Enhanced Learning Through Reduced Fear.} 
The stress and fear reduction had positive implications on students’ learning behavior and perception of testing. One student explained that humor \student{motivated me to try more edge cases because it felt less scary}{53-DE-F}. This connection between emotional state and exploration is important. When students felt less anxious about making mistakes, they became more willing to experiment, which is important in software testing. Another student described a similar effect on their focus: \student{Humor helped me worry less about what data are used to test and focus more on what is intended to be tested}{20-CA-M}. 

\emph{Context-Dependency and Neutral Response.}  
However, not all students found humor universally beneficial. Some raised concerns about context and personal preferences. One student worried that poorly executed humor could be \student{edgy, which would seem like trolling}{08-CA-M}, especially in courses that have otherwise low teaching quality. Another noted: \student{It was fun, but I can also see how people might be offended if its not for them}{29-CA-F}. Thus, humor requires an general awareness and reflection of the students: \student{Humor varies a lot between people. [...] There are many things we find funny and laugh about together, but that same humor can feel out of place, offensive to others. So this makes me wonder how humor translates into software testing}{14-CA-M}. 

\textbf{Comparison of Countries. }
\Cref{tab:country_comparison} shows no significant differences between students from the countries. This indicates that the emotional effects of humor in software testing might be similar across both educational contexts.

\textbf{Comparison of Genders. }
\Cref{tab:gender_comparison} shows statistically significant gender differences for more than half of the emotional engagement metrics.

We observed the largest effects for making testing more engaging and less monotonous, where female students reported stronger positive effects than male students (\emph{EM03}, \emph{EM04}). We observed medium-sized effects for enjoyment, reduced anxiety, and reduced intimidation, again with higher agreement among female students (\emph{EM02}, \emph{EM06}, and \emph{EM07}). 

Overall, these results suggest that humor supported emotional engagement for the vast majority of students, but especially for female students.

\begin{tcolorbox}[colback=blue!5, colframe=blue!50]
\textbf{RQ1.} Most students, particularly females, perceived humor as beneficial by increasing enjoyment and engagement, while reducing monotony.
\end{tcolorbox}

\subsection{RQ2: Humor on Students' Sense of Belonging}
\begin{figure}
	\centering
    \subfloat[\centering Retrospective perception of sense of belonging.]{{\includegraphics[width=0.8\linewidth]{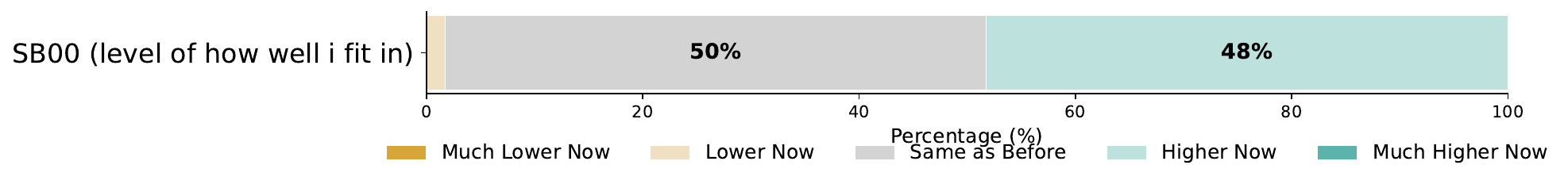} }}%
    \qquad
    \subfloat[\centering Perception of sense of belonging.]{{
	\includegraphics[width=0.8\linewidth]{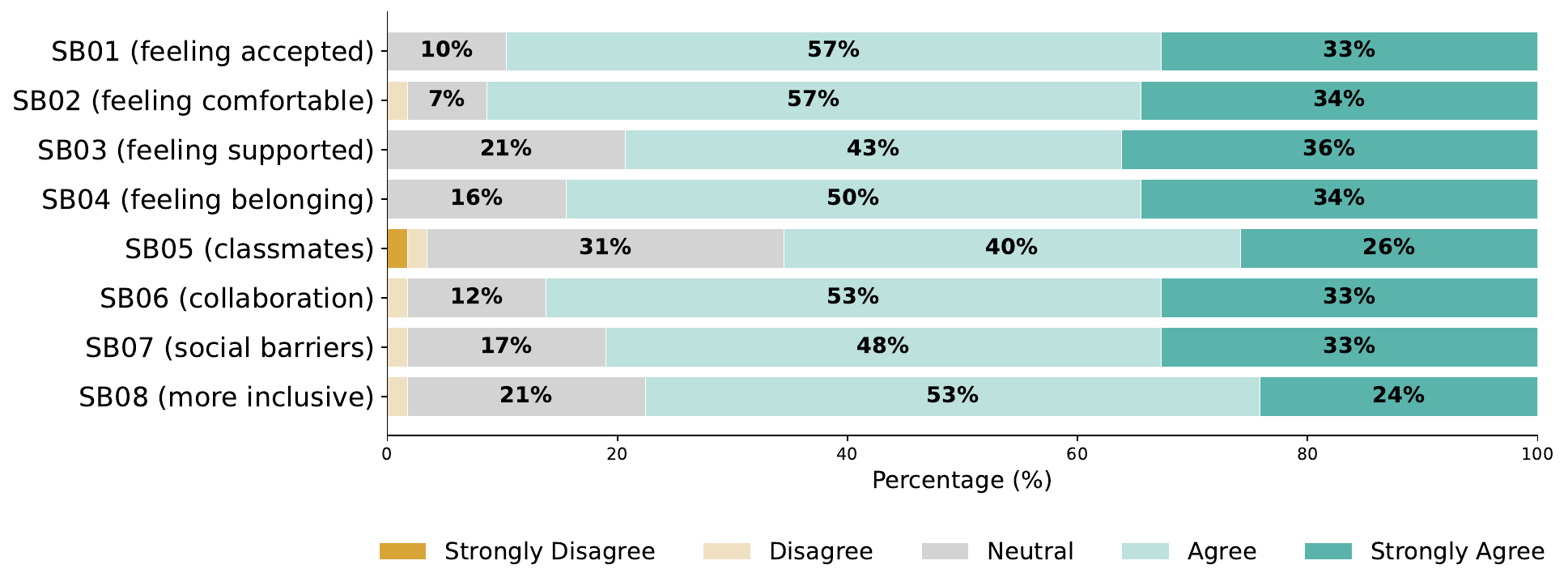}}}
	\caption{Students' perceived sense of belonging.}
	\label{fig:senseofbelonging}
\end{figure}
We asked students about their belonging and social experience in the classroom (\emph{SB00--SB09}).

\textbf{Overall Sense of Belonging. }
\Cref{fig:senseofbelonging} shows students' perceived sense of belonging. 
Across all items (\emph{SB00--SB08}), students reported high agreement on humor increasing their sense of belonging in software testing. 
Nearly half of the students reported a stronger sense of belonging compared to before the assignment (48.3\%), while half reported no change (50.0\%, \emph{SB00}).

Students strongly agreed that the humor within the assignment made them feel comfortable and accepted in class. The majority reported feeling comfortable during the assignment (91.4\%, 4.24, \emph{SB02}) and feeling accepted by others (89.7\%,  \emph{SB01}). Many students also reported that humor made collaboration more enjoyable (86.2\%, \emph{SB06}) and strengthened their feeling of belonging in the classroom (84.5\%, \emph{SB04}). In addition, students perceived humor as helpful for breaking down social barriers (81.0\%, \emph{SB07}).

We observed moderate agreement for items related to inclusion and social connection. Students reported that humor made testing more inclusive (77.6\%, \emph{SB08}) and helped them feel more connected to classmates (65.5\%, \emph{SB05}). A small number of students disagreed with this item, which suggests that social connection was not equally benefical for everyone.

\textbf{Comparison of Countries. }
According to \Cref{tab:country_comparison}, we observed two statistically significant differences between countries. Canadian students reported higher values for feeling supported (\emph{SB03}) and feeling a sense of belonging (\emph{SB04}) than German students, with small to medium effect sizes. Overall, Canadian students reported a slightly higher sense of belonging, which may be related to their longer exposure to the humor elements and opportunities for interaction during the intervention.

\textbf{Comparison of Genders. }
According to \Cref{tab:gender_comparison}, we observed statistically significant gender differences for almost all sense of belonging items. Female students consistently reported a higher sense of belonging than male students.

The strongest effects appeared for collaboration and inclusion. Female students reported that humor made collaboration more enjoyable (\emph{SB06}) and testing more inclusive (\emph{SB08}) to a greater extent than male students. We observed medium-sized effects for feeling accepted, comfortable, supported, connected, and overcoming social barriers (\emph{SB01--SB05}, \emph{SB07}).

The only item without a significant gender difference was the retrospective perception of fitting into software development (\emph{SB00}). This suggests that while humor influenced students’ classroom experience, a single assignment may not be sufficient to change students’ broader sense of belonging in software engineering.

\textbf{Qualitative Insights.} 
The qualitative analysis (\emph{SB09}; \Cref{tab:themes}) revealed that students perceived humor mainly as a mediator for social connection.

\emph{Foundation for Connection.} 
Students explained that humor created the emotional conditions necessary for social connection. The lightness and casualness introduced by humor reduced tension in the learning environment, which then made students more comfortable interacting with peers. 
This emotional foundation is important since it lowered the barriers to interaction due to the nature of humor: \student{humour makes me feel like I belong more in a team or in this field (Software engineering, Computer science...). Humour is friendly and informal so it's fun to encounter it in places where I might have a doubt on where I stand next to other people}{22-CA-M)}.

\emph{Communication.} 
Building on this relaxed foundation, humor provided students with a shared experience to discuss. Several students mentioned that humorous elements gave them something to talk about beyond technical topics. One student noted that humor \student{allowed to has software related talks with me peers without necessary talking about the software specifics}{21-CA-M)}. This is important because it suggests humor created common ground for students who might otherwise struggle to connect, especially early in a course. 
One student reflected: \student{I got along really well with my fellow students, which was actually the first time this semester, because I hadn't really dared to before.}{58-DE-F)}.

\emph{Community Building and Inclusiong}
For several students, these social interactions shown feelings of belonging within the groups, not being alone. One female student stated: \student{I usually feel like I don't fit in but this made me feel like part of the group}{31-CA-F.} This response is particularly important given ongoing concerns about inclusion in software engineering education. The humor helped increasing particpation by not only focusing on technical (prior) expertise, but connecting the students:
\student{It felt less strict and more like anyone could contribute, regardless of how good they were at testing or programming beforehand.}{56-DE-F}.

\begin{tcolorbox}[colback=blue!5, colframe=blue!50]
\textbf{RQ2.} Most students reported feeling comfortable and accepted, with female students reporting especially significant positive influences.
\end{tcolorbox}

\subsection{RQ3: Humor on Students' Creative Thinking}
\begin{figure}
	\centering
    \subfloat[\centering Retrospective perception of creativtiy.]{{\includegraphics[width=0.8\linewidth]{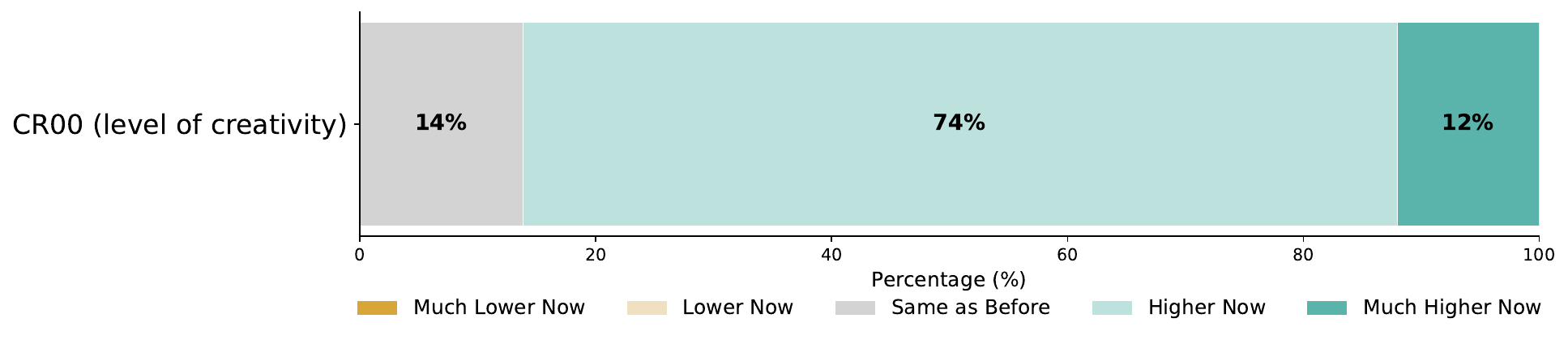} }}%
    \qquad
	\subfloat[\centering Perception of creativity.]{{\includegraphics[width=0.8\linewidth]{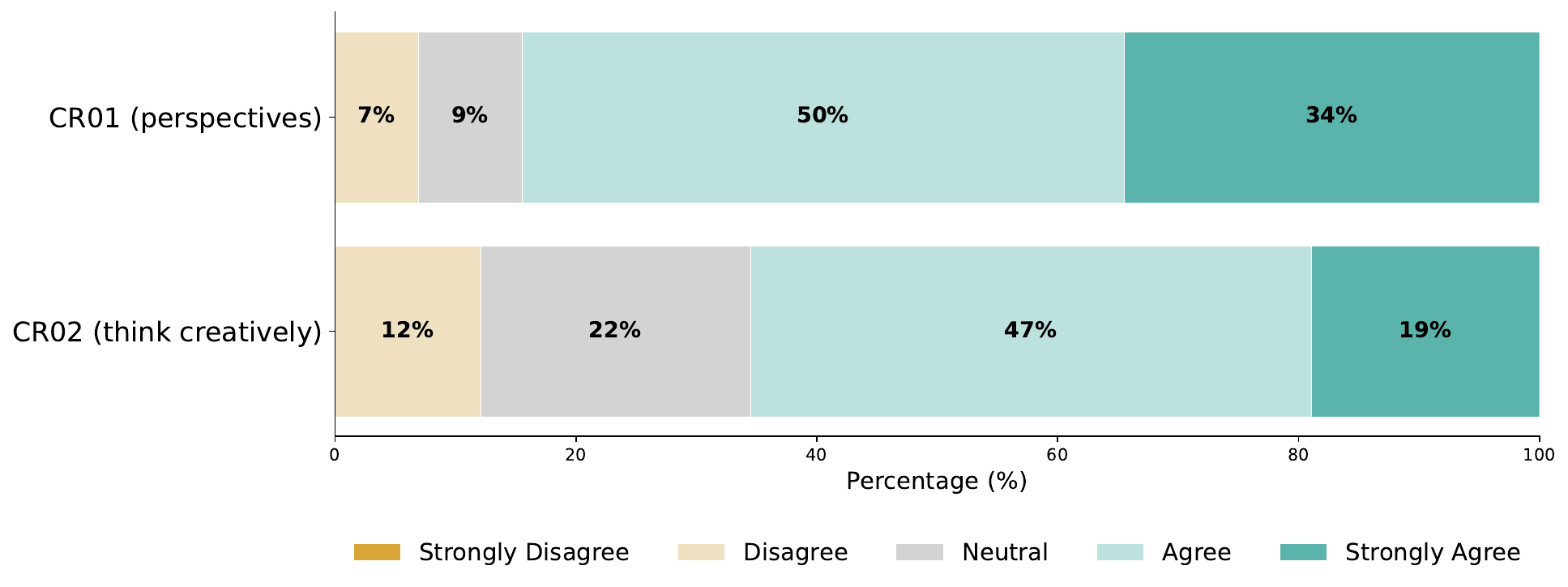} }}
	\caption{Students' perception of creativity.}
	\label{fig:creativity}
\end{figure}
We asked students about their creativity during testing (\emph{CR00--CR02}).

\textbf{Overall Creativity. }
\Cref{fig:creativity} shows the distrubtion of students' repsonses related to creative thinking in software testing. 
A large majority of students reported higher creativity after the assignment compared to before (86.2\%, \emph{CR00}).

Students reported particularly strong agreement that humor encouraged creative thinking in testing (84.5\%, \emph{CR01}). They also reported that humor helped them see testing tasks from different perspectives (65.5\%, \emph{CR02}). At the same time, a small group of students disagreed with this item (12 \%), which may indicate that some students focused more strongly on technical correctness than on creative exploration.

\textbf{Comparison of Countries. }
According to \Cref{tab:country_comparison}, we observed no statistically significant differences between the two countries. 
German students reported slightly higher retrospective creativity (\emph{CR00}), while Canadian students reported slightly higher agreement for encouraging creativity and seeing different perspectives (\emph{CR01--CR02}). This suggests that humor supports creative thinking across different educational contexts.

\textbf{Comparison of Genders. }
According to \Cref{tab:gender_comparison}, we observed statistically significant gender differences for all creativity items. 
Across all metrics, female students reported stronger creative benefits than male students, with medium to large effect sizes. This indicates that humor may be particularly effective in supporting creative engagement for female students in software testing.

\begin{tcolorbox}[colback=blue!5, colframe=blue!50]
\textbf{RQ3.} Most students perceived humor as supporting creative thinking in software testing, with female students reporting strong benefits.
\end{tcolorbox}

\section{Discussion and Future Work}

\emph{Synthesis of Results. }
Our results show that humor in software testing assignments is not mainly about reducing negative emotions such as frustration or stress. Instead, humor mainly creates a positive learning environment. It increases enjoyment and engagement in software testing, it improves the sense of belonging to a group that shares interests and cultural references, and it nurtures creativity when preparing the testing assignments. 

These findings suggest that humor in software testing works mainly as a social-emotiional mechanism, rather than as a frustration-reduction tool. 
Humor does not remove the challenges of software testing education, but it changes how students emotionally approach these challenges. Students still perceive testing as frustrating, but they experience it as more enjoyable and engaging. 
Humor also influences the social climate of the classroom, indicating that humor may be particularly effective in reducing social distance and supporting participation in software testing contexts. This sense of belonging is directly actionable when teaching the importance of collaboration and of complemeray skills when planning and doing software testing \cite{kanij2015empirical}.

From our qualitative findings, we see a pathway from humor to belonging as humor creates stress relief and lightness within the courses (RQ1), which fosters a relaxed environment where students feel comfortable expressing themselves socially (RQ2) and creatively (RQ3). This emotional safety enables communication and shared experiences, which in turn build community bonds. 

\emph{Cultural and Country Differences. }
Across RQ1 and RQ3, we found no statistically significant differences between countries, suggesting that the emotional and creative effects of humor in testing assignments are not strongly culture- or geographics-dependent. 
The only differences appeared for sense of belonging (RQ2), where Canadian students reported higher values. This suggests that the cultural or institutional context may influence how humor is incorporated and perceived in social support and belonging since the assignment for the Canadian students was much longer and the diverse and international student population may also have influenced how humor was received socially.

\emph{Gender Differences. }
We observed gender differences across all three research questions, with female students reporting stronger positive effects of humor. 
These differences were especially strong for sense of belonging. This pattern suggests that humor may help counteract aspects of software testing that female students otherwise perceive as unwelcoming, as one female student mentioned: \student{The whole environment was different, everyone was so relaxed and not this tension you usually have when being with boys}{13-CA-F}.

\subsection{Practical Implications and Future Work}
Our results indicate that humor is a beneficial addition to software testing education. Since the strongest benefits appear in our data when humor is used to shape the classroom climate, rather than to simplify or reduce technical demands, we encourage educators to use humor to improve the learning atmosphere in testing courses, especially to support social interaction and participation. Our study shows that this introduction of humor in software tesitng education can take multiple forms. First,  humor can be introduced on different time scales,  from a single,  three hour lecture or assignment, as seen in the German students’ responses, to a longer term presence of humor in lectures and assignments all along the semester, as seen in Canada. Second, it humor can be introduced through different pedagogical tools, such as the professor using humorous examples to illustrate fundamental testing concepts, to the students practicing humor as part of their testing assignment.
Future studies should examine how humor influences \emph{cognitive} engagement and well-being in testing assignments, and compare these effects with findings from broader computing education research~\cite{erdogdu2021a}. 

\section{Conclusions}
This empirical study showed that humor can be a meaningful practice in software testing education. Across two educational contexts, humor consistently functioned as a social-emotional practice, improving how students emotionally engage with testing, relate to peers, and approach testing tasks creatively. In particular, female students experienced significant benefits in belonging, collaboration, and creativity, indicating that humor may help soften social barriers in male-dominated software engineering settings. Although humor alone is not a solution to inequities, it might offer a simple and accessible way to support more inclusive learning environments in software testing.

\section*{Acknowledgements}
The authors thank the teaching assistants for their support in explaining and assessing the humorous elements of the assignments, as well as all the students who took the courses and participated in this study. This work is supported by NSERC and by IVADO and the Canada First Research Excellence Fund. 

\bibliographystyle{IEEEtran}
\bibliography{references-humor}

\end{document}